\documentclass{ws-procs9x6}

\let\draftnote\relax
\let\trimmarks\relax

 \catcode`@=11
\def\@cite#1#2{\unskip\nobreak\relax
    \def\@tempa{$\m@th^{\hbox{\the\scriptfont0 #1}}$}%
    \futurelet\@tempc\@citexx}
\def\@citexx{\ifx.\@tempc\let\@tempd=\@citepunct\else
    \ifx,\@tempc\let\@tempd=\@citepunct\else
    \ifx;\@tempc\let\@tempd=\@citepunct\else
    \let\@tempd=\@tempa\fi\fi\fi\@tempd}
\def\@citepunct{\@tempc\edef\@sf{\spacefactor=\the\spacefactor\relax}\@tempa
    \@sf\@gobble}
\catcode`@=12

\def\lc{\lowercase}

\begin{document}

\title{
\hbox to \hsize{{\bf U\lc{niversity of} W\lc{isconsin} - M\lc{adison}}
\hfill\vtop{
\hbox{\bf MADPH-03-1320}
\hbox{\small J\lc{anuary} 2003}
\hbox{\hfil}}}
MULTI-MESSENGER ASTRONOMY: COSMIC RAYS, GAMMA-RAYS AND NEUTRINOS}

\author{\vspace*{-.35in}FRANCIS~HALZEN}

\address{Department of Physics, University of Wisconsin, Madison, WI
53706, USA}

\maketitle

\abstracts{Although cosmic rays were discovered a century ago, we do not know where or how they are accelerated. There is a realistic hope that the oldest problem in astronomy will be solved soon by ambitious experimentation: air shower arrays of 10,000 kilometer-square area, arrays of air Cerenkov telescopes and kilometer-scale neutrino observatories. Their predecessors are producing science. We will review the highlights:
\begin{itemize}
\item Cosmic rays: the highest energy particles and the GZK cutoff, the search for cosmic accelerators and the the Cygnus region, top-down mechanisms: photons versus protons?
\item TeV-energy gamma rays: blazars, how molecular clouds may have revealed proton beams, first hints of the diffuse infrared background?
\item Neutrinos: first results and proof of concept for technologies to construct kilometer-scale observatories.
\end{itemize}
}

\section{The High Energy Universe}

Very ambitious projects have been launched to extend conventional astronomy beyond wavelengths of $10^{-14}$\,cm, or GeV photon energy; see Fig.\,1. Besides gamma rays, protons (nuclei), neutrinos and gravitational waves are explored as astronomical messengers probing the high energy universe. The challenges are considerable:
\begin{itemize}
\item Protons are relatively abundant, but their arrival directions have been scrambled by magnetic fields.
\item $\gamma$-rays do point back to their sources, but are absorbed on extragalactic backgrounds at TeV- energy and above.
\item neutrinos propagate unabsorbed and without deflection throughout the universe but are difficult to detect.
\end{itemize}
Therefore, multi-messenger astronomy may not just be an advantage, it may be a necessity for solving some of the outstanding problems of astronomy at the highest energies such as the identification of the sources of the cosmic rays, the nature of the initial event(s) triggering gamma ray bursts and the particle nature of the dark matter. 

% figure 1
\begin{figure}[t]
\centering\leavevmode
\includegraphics[width=3.75in,angle=-90]{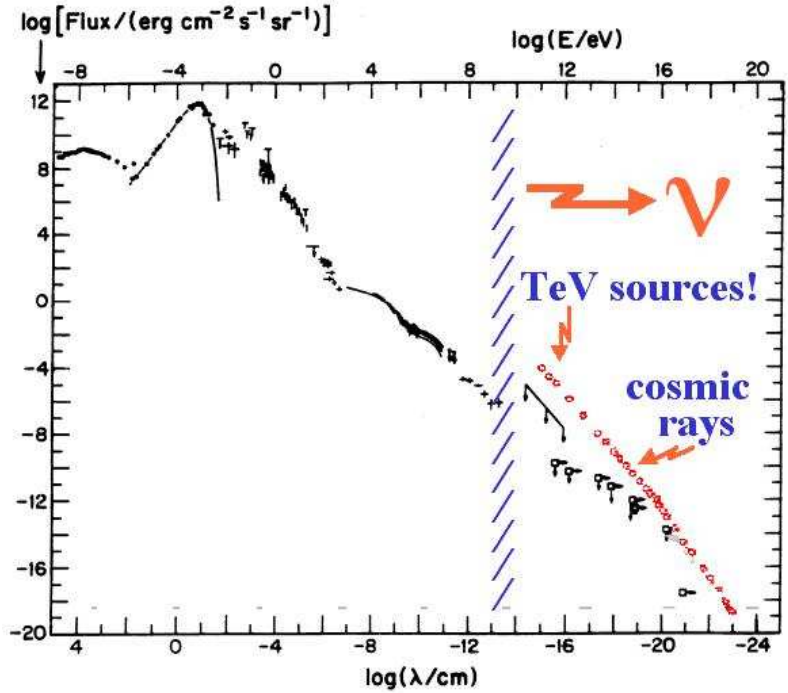}
\caption[]{The diffuse flux of photons in the universe, from radio waves to GeV-photons\cite{turner}. Above tens of GeV only limits are reported although individual sources emitting TeV gamma-rays have been identified. Above GeV energy cosmic rays dominate the universal diffuse spectrum.}
\end{figure}

\section{Cosmic Rays: Protons\dots Fe}

The flux of cosmic rays is summarized in Fig.\,2a,b\cite{gaisseramsterdam}. The cosmic ray spectrum peaks in the vicinity of 1\,GeV; below this energy the particles are screened by the sun, above the spectrum follows a broken power law. The two power laws are separated by a feature referred to as the ``knee"; see Fig.\,2a. There is evidence that the cosmic rays, up to at least several EeV, originate in galactic sources. This correlation disappears in the vicinity of a second feature in the spectrum dubbed the ``ankle". Above the ankle the gyroradius of a proton exceeds the size of the galaxy and the generally accepted assumption is that we are  witnessing the onset of an extragalactic component in the spectrum that extends to energies beyond 100\,EeV.

%% fig.2
\begin{figure}[t]
\centering\leavevmode
\includegraphics[width=4.5in]{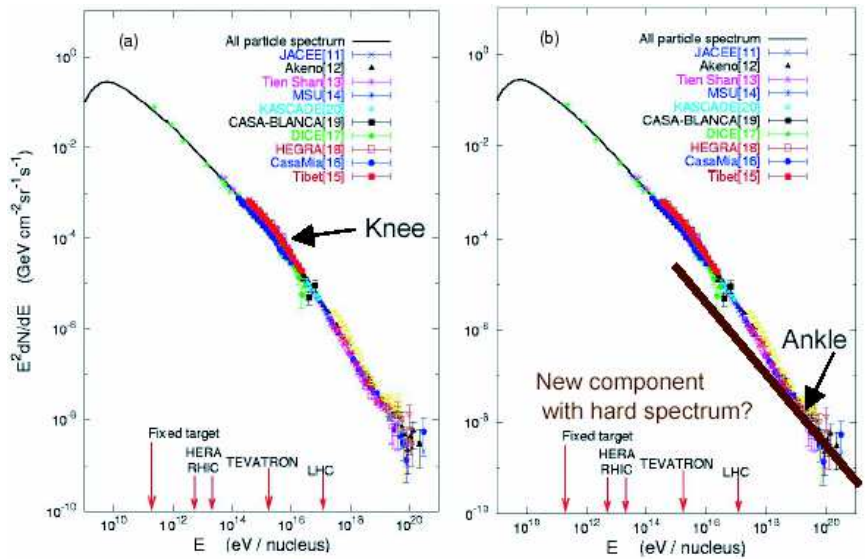}
\caption[]{At the energies of interest here, the cosmic ray spectrum consists of a sequence of 3 power laws. The first two are separated by the ``knee" (left panel), the second and third by the ``ankle". There is evidence that the cosmic rays beyond the ankle are a new population of particles produced in extragalactic sources; see right panel\cite{gaisseramsterdam}.}
\end{figure}

The principal reason why this impressive set of data fails to reveal the origin of the particles is undoubtedly that their directions have been scrambled by the microgauss galactic magnetic fields. At EeV energies ``proton astronomy'' may however be possible because the arrival directions of electrically charged cosmic rays are no longer scrambled by the ambient magnetic field of our own galaxy. Protons point back to their sources with an accuracy determined by their gyroradius in the intergalactic magnetic field~$B$:
\begin{equation}
\theta \cong {d\over R_{\rm gyro}} = {dB\over E} \,,
\end{equation}
where $d$ is the distance to the source. Scaled to units relevant to the problem,
\begin{equation}
{\theta\over0.1^\circ} \cong { \left( d\over 1{\rm\ Mpc} \right)
\left( B\over 10^{-9}{\rm\,G} \right) \over \left( E\over
3\times10^{20}\rm\, eV\right) }\,.
\end{equation}
Speculations on the strength of the inter-galactic magnetic field range from $10^{-7}$ to $10^{-12}$~Gauss. For the distance to a nearby galaxy at 100~Mpc, the resolution may therefore be anywhere from sub-degree to nonexistent. It is still reasonable to expect that the arrival directions of the very highest energy cosmic rays may provide information on the location of their sources. Conversely, they may yield indirect information on intergalactic magnetic fields. Determining their strength by conventional astronomical means has turned out to be challenging.

\subsection{The Highest Energies: Conflicting Signals}

We first concentrate on the extra-galactic component of the flux shown in Fig.\,2b. Where experiments are concerned, all signatures for the particle nature of these cosmic rays converge on only one possible conclusion: they are protons or, possibly, nuclei. Just like the universe is partially obscured for photons with energy in excess of tens of TeV because they annihilate into electron pairs in interactions with the infrared photon background, similarly protons  interact with background light, predominantly by photoproduction of the $\Delta$-resonance with CMB photons, $p + \gamma_{CMB} \rightarrow \Delta \rightarrow \pi + p$, above a threshold energy $E_p$ of roughly 50\,EeV given by:
\begin{equation}
2E_p\epsilon > \left(m_\Delta^2 - m_p^2\right) \,;
\label{eq:threshold}
\end{equation}
where $\epsilon$ is the energy of the CMB photon. The universe is, therefore, also opaque to the highest energy cosmic rays, with an absorption length~of
\begin{eqnarray}
\lambda_{\gamma p} &=& (n_{\rm CMB} \, \sigma_{p+\gamma_{\rm
CMB}})^{-1}\cong10\rm\ Mpc
\end{eqnarray}
when their energy exceeds 50\,EeV. This so-called GZK cutoff establishes a universal upper limit on the energy of the cosmic rays. The cutoff is robust, depending only on two known numbers: $n_{\mathrm{CMB}} = 400\rm\,cm^{-3}$ and $\sigma_{p+\gamma_{\mathrm{CMB}}} = 10^{-28}\rm\,cm^2$.

Cosmic rays do reach us with energies exceeding 100\,EeV. In October 1991, the Fly's Eye cosmic ray detector recorded an event of energy $3.0\pm^{0.36}_{0.54}\times 10^{20}$\,eV.\cite{flyes} This event, together with an event recorded by the Yakutsk air shower array in May 1989,\cite{yakutsk} of estimated energy $\sim\,2\times10^{20}$\,eV, constituted at the time the two highest energy cosmic rays ever seen. Their energy corresponds to a center of mass energy of the order of 700~TeV or ${\sim}50$ Joules, almost 50 times LHC energy. In fact, all experiments\cite{web} have detected cosmic rays in the vicinity of 100~EeV since their discovery by the Haverah Park air shower array.\cite{WatsonZas} The AGASA air shower array in Japan\cite{agasa} has by now accumulated an impressive 10 events with energy in excess of $10^{20}$\,eV.\cite{ICRC}

%% fig 3 orig position

With a particle flux of order 1 event per km$^2$ per century, these events can only be detected by using the earth's atmosphere as a particle detector. The experimental signature is a shower initiated by the particle in the upper atmosphere. The primary particle creates an electromagnetic and hadronic cascade.  The shower grows to a shower maximum and is subsequently absorbed by the atmosphere. The shower can be observed by: i) sampling the electromagnetic and hadronic components when they reach the ground with an array of particle detectors such as scintillators, ii) detecting from the ground the fluorescent light emitted by the nitrogen atoms excited by the passage of the shower particles through the atmosphere, iii) detecting from the ground the Cerenkov light emitted by shower particles high in the atmosphere, and iv)~detecting muons and neutrinos produced in the hadronic component of the air shower.

The energy resolution of these experiments is a critical issue. Several experiments using the first two techniques agree on the energy of EeV-showers within a resolution of $\sim$\,25\%. Additionally, there is a systematic error of order 10\% associated with the modeling of the showers. All techniques are indeed subject to uncertainties associated with particle simulations that involve physics beyond the LHC. If the final outcome turns out to be an erroneous inference of the energy of the shower because of new physics associated with particle interactions at the $\Lambda_{\mathrm{QCD}}$ scale, we will have to contemplate that discovery instead.

%% fig3
\begin{figure}
\centering\leavevmode
\includegraphics[width=3.75in]{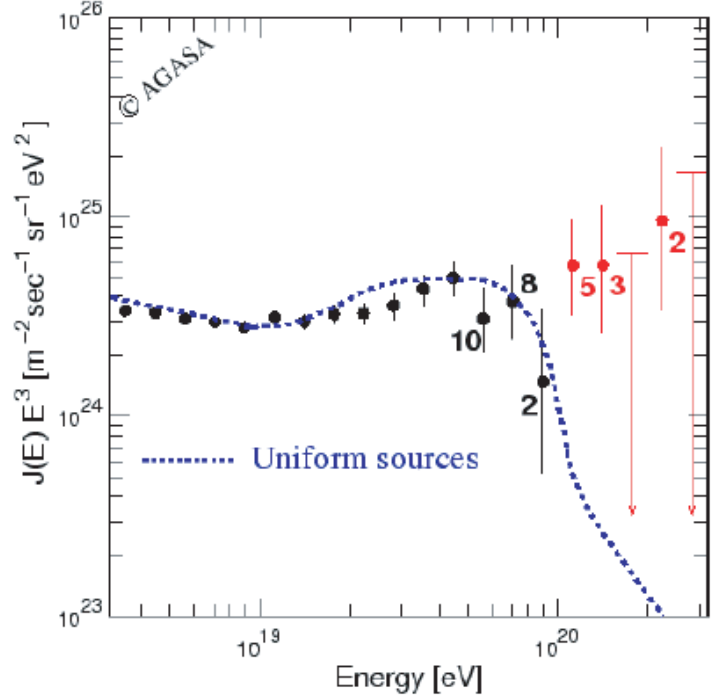}
\caption[]{The flux of the highest energy cosmic rays measured by the AGASA experiment is confronted with the assumption that they are produced in sources that are uniformly distributed through the universe. The cutoff in energy resulting from the interaction of the cosmic rays with the microwave background is not observed}
\end{figure}

HiRes and AGASA agree that cosmic rays with energy in excess of 10\,EeV are not a feature of our galaxy and that their spectrum extends beyond 100\,EeV. They disagree on everything else. The particle fluxes measured by the two experiments are shown in Figs.\,3 and 4 along with rather contrasting interpretations of the observations. The AGASA results in Fig.\,3 are fitted to a power-law spectrum produced by sources uniformly distributed through the universe. The fit inevitably shows the GZK cutoff which their data does not exhibit; the disagreement is at the 4\,$\sigma$ level and is produced by 10 excess events. In contrast, in Fig.\,4 a similar fit does accommodate the HiRes data, including the highest energy event\cite{sokolsky}. Agreement is achieved by assuming that the cosmic ray sources have cosmologically evolved with redshift with a dependence of $(1+z)^3$. Because of the limited statistics, interpreting the measured fluxes in Figs.\,3 and 4 as a function of energy is like reading tea leaves; one cannot help however reading different messages in the two spectra. The conflict is aggravated by the fact that the AGASA data show evidence at the 5\,$\sigma$-level that the highest energy cosmic rays come from point sources; see Fig.\,5. The HiRes data do not support this conclusion.

%% fig4
\begin{figure}
\centering\leavevmode
\includegraphics[width=3.75in,angle=-90]{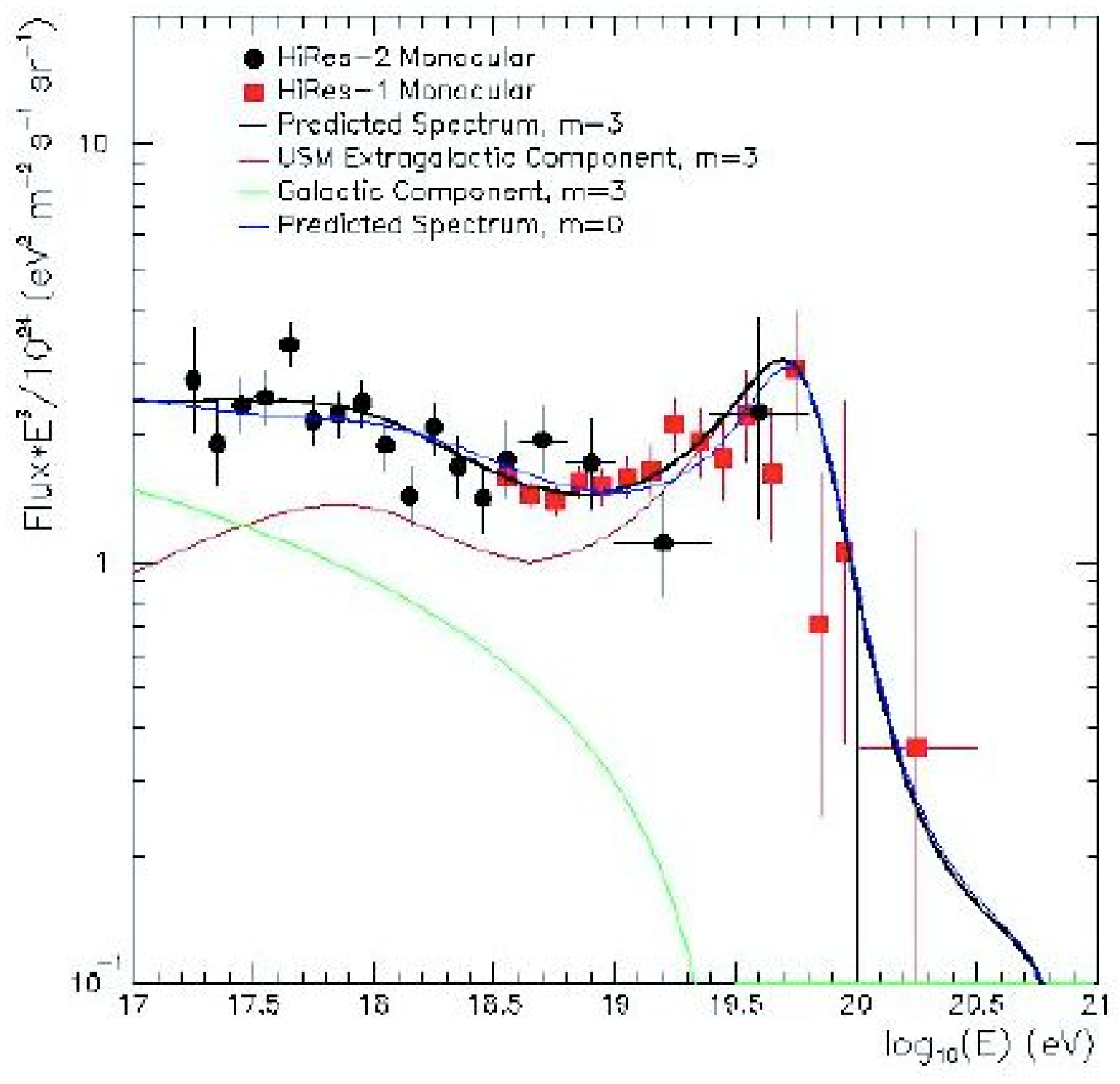}
\caption[]{Same as Fig.\,3 for the HiRes experiment. The final fit assumes that the sources evolved with redshift following a $(1+z)^3$ dependence.}
\end{figure}

%% fig5
\begin{figure}[t]
\includegraphics[width=4.5in]{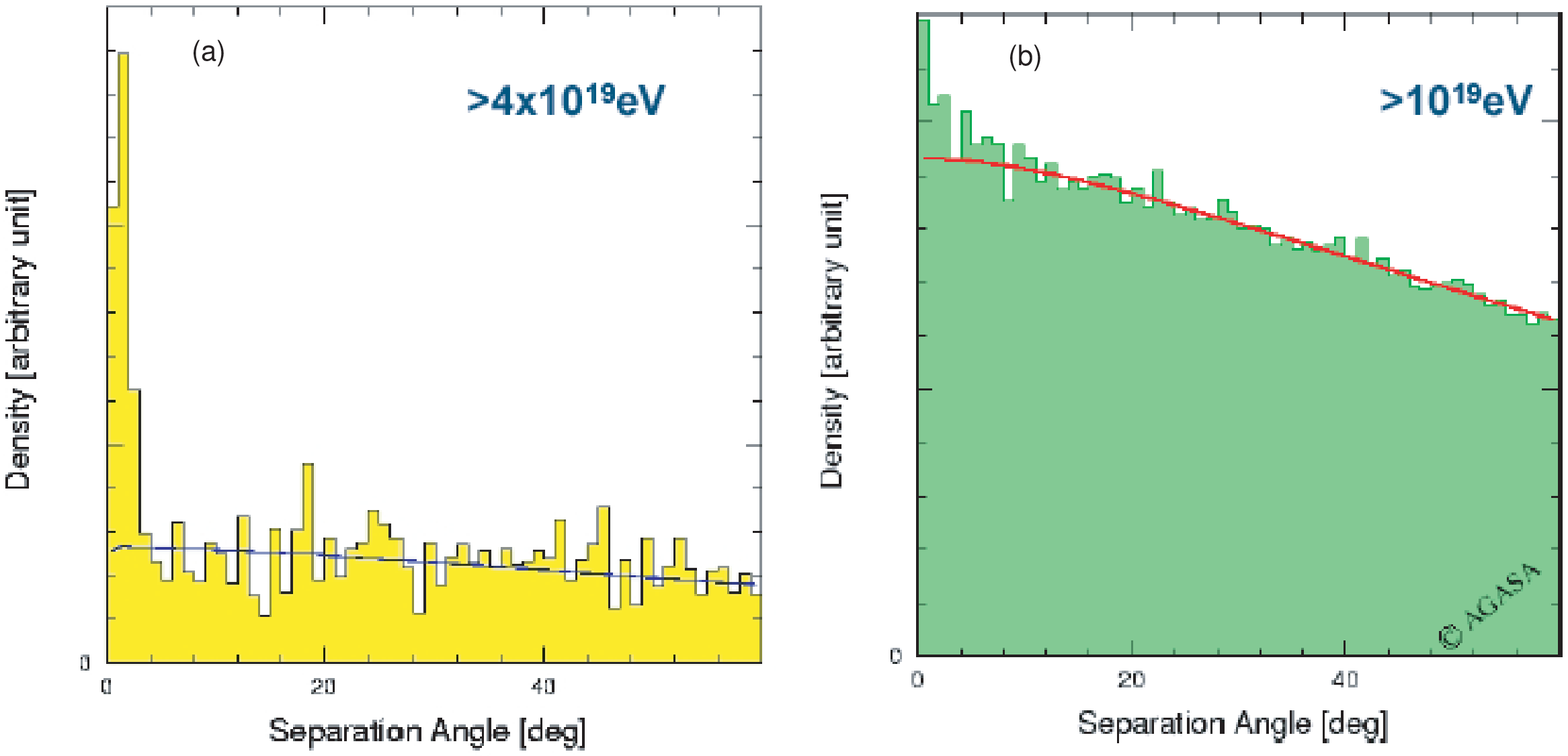}
\caption[]{Plotting the distribution of angles between all pairs of cosmic rays, the AGASA experiment reveals a correlation (the peak at zero angle in the left panel) that diminishes with decreasing energy (right panel).}
\end{figure}

If the AGASA results were to be confirmed, we would be left with few options: i)~the highest energy cosmic rays are accelerated in nearby sources, ii)~they do reach us from distant sources that accelerate cosmic rays to even higher initial energies than observed, thus exacerbating the acceleration problem, or iii)~the highest energy cosmic rays are not protons, a possibility not favored by data as we will discuss further on. The first possibility raises the considerable challenge of finding an appropriate accelerator in our local cluster. It is not impossible that all cosmic rays are produced by the active galaxy M87\cite{farrar} or by a relatively nearby gamma ray burst which exploded a few hundred years ago. Stecker \cite{stecker2} has speculated that the highest energy cosmic rays may be Fe nuclei with a delayed GZK cutoff. The details are complicated but the relevant quantity in the problem is $\gamma=E/Am_p$, where $A$ is the atomic number. For a fixed observed energy, the smallest boost towards GZK threshold is associated with the largest atomic mass, {\it i.e.}~Fe.

\subsection{Direct Evidence for Cosmic Accelerators?}

Despite impressive progress in our understanding of how charged particles can be efficiently accelerated in shocks\cite{shocks}, for instance those initiated by supernova explosions or gamma ray bursts, the ``evidence" that these may be the sources of the cosmic rays is limited to simple dimensional analysis and energetics --- no cosmic ray accelerator has been conclusively identified.

First dimensional analysis. It is sensible to assume that, in order to accelerate a proton to energy $E$ in a magnetic field $B$, the size $R$ of the accelerator must be larger than the gyroradius of the particle:
\begin{equation}
R > R_{\rm gyro} = {E\over B}\,,
\end{equation}
{\it i.e.} the accelerating magnetic field must contain the particle orbit. This condition yields a maximum energy
\begin{equation}
E = \Gamma BR
\end{equation}
by dimensional analysis and nothing more. The $\Gamma$-factor has been included to allow for the possibility that we may not be at rest in the frame of the cosmic accelerator resulting in the observation of boosted particle energies. For nuclei with charge $Ze$, the maximal energy can be raised by a factor $Z$. This is the famous Hillas formula.

Theorists' imagination regarding the accelerators is limited to dense regions where exceptional gravitational forces create relativistic particle flows of charged particles: the dense cores of exploding stars, inflows on supermassive black holes at the centers of active galaxies, annihilating black holes or neutron stars? All speculations involve collapsed objects and we can therefore replace $R$ by the Schwartzschild radius
\begin{equation}
R \sim GM/c^2
\end{equation}
to obtain
\begin{equation}
E \sim \Gamma BM \,.
\end{equation}
It may not be a coincidence that for a solar mass black hole and fields of order $10^{12}$\,Gauss, energies up to 10 EeV can be reached for $\Gamma=1$. This has led to the speculation that solar mass collapsed objects associated with galactic supernovae, pulsars and mini-quasars and the like, may be the cosmic accelerators responsible for the flux in Fig.\,2a. These speculations are further supported by the fact that the energy injected by supernovae exploding at the rate of a few per century can supply the energy density in galactic cosmic rays indicated by the data in Fig.\,2a.

Galactic magnetic fields are not sufficiently strong and there are no galactic sources sufficiently large or massive to reach the energies of the highest energy cosmic rays in Fig.\,2b, even by dimensional analysis. This limits their sources to extragalactic objects. Active galaxies powered by a billion solar mass black holes are candidates. With kilo-Gauss fields we reach 100\,EeV using Eq.\,(6). The jets emitted by the central black hole in blazars could achieve similar energies in shocks boosted in our direction by a $\Gamma$-factor of 10, possibly higher. The neutron star or black hole remnant of a collapsing supermassive star could support magnetic fields of $10^{12}$\,Gauss, possibly larger. Shocks with $\Gamma > 10^2$ emanating from the collapsed black hole could be the origin of gamma ray bursts and, possibly, the source of the highest energy cosmic rays. Gamma ray bursts are not only the dimensional winner --- much is made from the fact that their total injection rate in the universe at a rate of a few solar masses per day matches the energy density of cosmic rays beyond the ``ankle" in Fig.2b\cite{vietriwaxman}. On the basis of dimensional and energetic arguments, supernova remnants and gamma ray bursts have emerged as the leading speculations for the sources of the cosmic rays.

On the observational front, no experiment has produced conclusive evidence for a cosmic accelerator, neither by observation of a point source nor by detecting gamma rays or neutrinos from the decay of pions associated with the accelerator's beam. There may be some hints:
\begin{enumerate}
\item With all the controversy over the GZK cutoff above 50\,EeV --- more about that later --- one may have missed the fact that below 10 EeV the AGASA air shower experiment has revealed a correlation of the arrival direction of the cosmic rays to the Cygnus region at the 4\,$\sigma$ level\cite{agasa}. At EeV energy, neutrons can reach us before decay to form a cluster of pointing events from source(s) concentrated in the spiral arm of our galaxy in the Cygnus direction. The particles could also be photons as their EeV energies exceed the PeV energy region where the galaxy is highly obscured by pair production on microwave photons. Previously, the highest fluctuation in a map of cosmic ray arrival directions observed by the Kiel air shower experiment pointed in the Cygnus direction\cite{kiel}, a possibly valid clue that may have been lost in unsuccessful attempts to identify the source with  Cygnus X-3. Finally, the HEGRA experiment has detected an extended TeV $\gamma$-ray source in the Cygnus region with no clear counterpart and a spectrum not easily accommodated with synchrotoron radiation by electrons\cite{hegracygnus}. The discovery of a source with no counterpart is a first for air Cerenkov telescopes. The model proposed is that of a proton beam, accelerated by a nearby
mini-quasar or possibly Cygnus X-3, interacting with a molecular cloud to produce pions that are the source of the gamma rays.
\item The unusual spectrum of TeV $\gamma$-rays emitted by a molecular cloud near SNR RX J1713.7-3946 has been interpreted as originating from the decay of neutral pions produced by high energy protons accelerated by the supernova remnant\cite{cangoroo}. The target cloud is a known diffuse ASCA X-ray source. It has been argued that the very hard spectrum in the 1--10\,KeV range is produced by bremsstrahlung of protons\cite{aharonian}. With an energy content in relativistic protons of order a few percent of what is inferred from the X-ray observation for low energy protons, one can accommodate the CANGOROO observation at TeV energy.
\item Suspects for cosmic ray acceleration also include SN1006, Sgr A East and Cass A\cite{crocker}.
\end{enumerate}

These hints for cosmic accelerators, though hardly conclusive, are certainly worth pursuing. Using these observation in conjunction with simple energetics and the equality of neutral and charged pions produced, it has been pointed out\cite{alvarezhalzen} that a beamdump associated with  a molecular cloud should produce TeV-energy neutrinos at a rate of 10\,km$^{-2}\rm\,year^{-1}$. Their observation would produce uncontrovertible evidence for a cosmic ray accelerator. This is also one more indication that it takes kilometer-scale neutrino observatories to detect neutrinos associated with the high energy cosmic rays\cite{waxmanbahcall}; more on this later. 
   
\subsection{The Highest Energies: Bottom-Up or Top-Down?}

The astrophysics problem of accelerating cosmic rays to EeV-energy and beyond is so daunting that many believe that the highest energy cosmic rays are the decay products of remnant particles or topological structures created in the early universe\cite{berezinski}. In these scenarios the highest energy cosmic rays are predominantly photons. A topological defect from a phase transition in a Grand Unified Theory with typical energy scale of $10^{24}$\,eV will suffer a chain decay into GUT particles X and Y that subsequently decay to familiar weak bosons, leptons and quark or gluon jets. Cosmic rays are in the end the fragmentation products of these jets. We know from accelerator studies that, among the fragmentation products of jets, neutral pions (decaying into photons) dominate, in number, protons by close to two orders of magnitude. Therefore, if the decay of topological defects is the source of the highest energy cosmic rays, they must be photons. This is a problem because there is evidence that the highest energy cosmic rays are not photons:

\begin{enumerate}
\item The highest energy event observed by the Fly's Eye is not likely to
be a photon \cite{vazquez}.  A photon of 300\,EeV will interact with the
magnetic field of the earth far above the atmosphere and disintegrate
into lower energy cascades --- roughly ten at this particular energy.
The detector subsequently collects light produced by the fluorescence of
atmospheric nitrogen along the path of the high-energy showers
traversing the atmosphere. The atmospheric shower profile of a 300\,EeV
photon after fragmentation in the earthÕs magnetic field is shown in
Fig.\,6. It disagrees with the data. The observed shower
profile does fit that of a primary proton, and,
possibly, a nucleus. The shower profile information is
sufficient,
however, to conclude that the
event is unlikely to be of photon origin. More precisely, it takes eight gamma ray showers to produce the observed shower by fluctuations.
\item The same conclusion is
reached for the Yakutsk event that is characterized by a very large number
of secondary muons from pion decay in a hadronic shower, inconsistent with a purely electromagnetic cascade
initiated by a gamma ray.
\item The AGASA collaboration claims evidence
for ``point" sources above 10\,EeV. The arrival directions are
however smeared out in a way consistent with primaries deflected by
the galactic magnetic field. Again, this indicates charged primaries
and excludes photons.
\item Finally, a recent reanalysis of the Haverah Park disfavors photon
origin of the primaries \cite{WatsonZas}.
\end{enumerate}

A possible way out of this dilemma is to identify the observed cosmic rays with the suppressed proton flux produced by top-down models, arguing that the dominant photon flux is efficiently absorbed by pair production on the diffuse radio background. This feat can be realized within the large ambiguities on the magnitude of the universal flux at radio-wavelengths shown in Fig.\,1. Doing this one inevitably boosts the primary flux of topological defects by more than one order of magnitude exacerbating the already questionable energy requirements of these models, but also augmenting the predicted neutrino flux to a level where it should be observed by AMANDA, and certainly by AMANDA\,II\cite{hoopertalk,3russians}.

%figure 6, chalonge figure 4
\begin{figure}[t]
\centering\leavevmode
\includegraphics[width=4.25in]{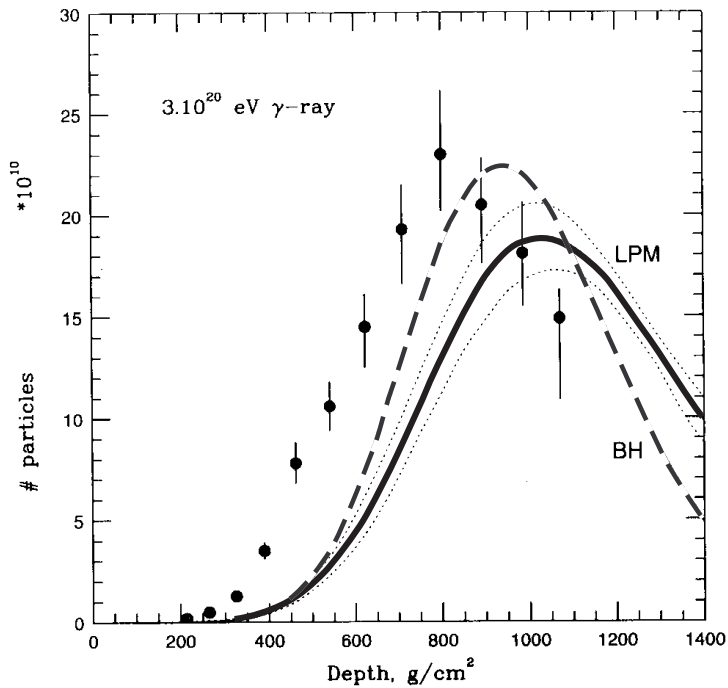}

\caption[]{The composite atmospheric shower profile of a $3\times
10^{20}$\,eV gamma ray shower calculated with Landau-Pomeranchuk-Migdal
(dashed) and Bethe-Heitler (solid)
electromagnetic cross sections. The central line shows the average shower
profile and the upper and lower lines show 1~$\sigma$ deviations --- not
visible for the BH case, where lines overlap. The experimental shower
profile is shown with the data points. It does not fit the profile of a photon
shower.}
\end{figure}

Could the primaries be neutrinos, even more abundantly produced by topological defects than photons? Standard model neutrino
physics is understood, even for EeV energies, and excludes neutrino cross sections that are sufficiently large to produce air showers. The average $x$ of the parton mediating the neutrino interaction is of order $x \sim \sqrt{M_W^2/s} \sim 10^{-6}$ so that the perturbative result for the neutrino-nucleus cross section is calculable from measured HERA structure functions. Because $Q^2\sim{M_W}^2$, even at 100\,EeV a reliable value of the cross section can be obtained based on QCD-inspired extrapolations of the structure function. The neutrino cross section is known to better than an order of magnitude. It falls 5 orders of magnitude short of the strong cross sections required to make a neutrino interact in the upper atmosphere to create an air shower.

Could EeV neutrinos be strongly interacting because of new physics? In theories with TeV-scale gravity, one can imagine that graviton exchange dominates all interactions, thus erasing the difference between quarks and neutrinos at the energies under consideration. The actual models performing this feat require a fast turn-on of the cross section with energy that (arguably) violates $S$-wave unitarity
\cite{han}.

All this leaves us with the reasonable conclusion that the highest energy cosmic rays are protons, or nuclei, as indeed indicated by all data.

\subsection{Technology to the Rescue: HiRes and Auger}

Given these mixed observational messages, speculating on the nature of the sources is essentially impossible, especially where the highest energies are concerned. We clearly need higher statistics experiments with good control over the difficult systematics. Over the next years, a qualitative improvement can be expected from the operation of the HiRes fluorescence detector in Utah. With improved instrumentation yielding high quality data from 2 detectors operated in coincidence, the systematics associated with the interplay between sky transparency and energy measurement can be studied in detail. The data shown in Fig.\,4 were taken in monocular mode. The Auger air shower array, under construction, is confronting the low rate problem by instrumenting a huge collection area covering 3000 square kilometers on an elevated plain in Western Argentina\cite{auger}. The elements of the array are 1600 water Cerenkov detectors spaced by 1.5\,km. For calibration, about 15 percent of the showers occurring at night will be viewed by 3 HiRes-style fluorescence detectors. The detector will observe several thousand events per year above 10\,EeV and tens above 100\,EeV, with the exact numbers depending on the detailed shape of the observed spectrum which is at present a matter of speculation.

\section{Gamma-Rays: tens of GeV to tens of TeV}

There are alternative ways to search for the sources of the cosmic rays. We anticipate indeed that secondary photons and neutrinos are associated with the highest energy cosmic rays, as was the case for the cosmological remnants previously discussed. These point to the sources!  The cartoon in Fig.\,7 illustrates why cosmic accelerators are also cosmic beam dumps that produce secondary photon and neutrino beams. Accelerating particles to TeV energy and above requires relativistic, massive bulk flows of charged particles. These are likely to originate from the exceptional gravitational forces associated with black holes or neutron stars. Accelerated particles therefore pass through intense radiation fields or dense clouds of gas surrounding the black hole leading to the production of secondary pions. These subsequently decay into photons and neutrinos that accompany the primary cosmic ray beam. Examples of targets for pion production include the external photon clouds that surround the central black hole of active galaxies, and the matter falling into the collapsed core of a dying supermassive star producing a gamma ray burst. Also, outside the sources, high energy particles produce secondary photons and neutrinos in interactions with molecular clouds near the accelerator, as previously discussed, and in collisions with microwave photons. In all examples, the target material, whether a gas of particles or of photons, is likely to be sufficiently tenuous for the primary proton beam and the secondary photon beam to be only partially attenuated. However, shrouded sources from which only neutrinos can emerge, as is the case for terrestrial beam dumps at CERN and Fermilab, are also a possibility.

%figure 7
\begin{figure}[t]
\centering\leavevmode
\includegraphics[width=3.5in]{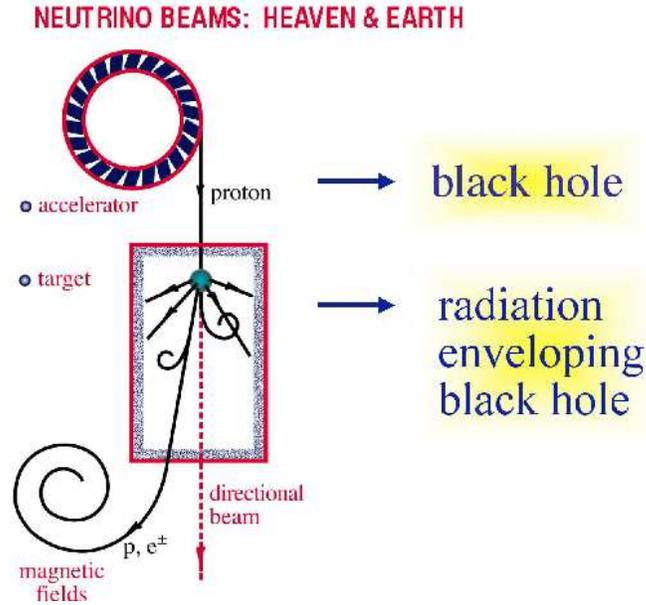}
\caption[]{Diagram of cosmic accelerator and beam dump.  See text for
discussion.}
\end{figure}

Ground-based gamma ray astronomy has become a mature science.\cite{weekes} A large mirror, viewed by an array of photomultipliers, collects the Cerenkov light emitted by shower particles in the high atmosphere and images the showers in order to determine the arrival direction as well as the nature of the primary particle. The great advantage is that a relatively modest telescope samples showers over a 40,000\,m$^2$ area at a height of approximately 10\,km. These experiments have opened a new window in astronomy by finding isolated sources with a photon spectrum extending to 20\,TeV, possibly beyond.

Observations have revealed TeV-emission from galactic supernova remnants and nearby quasars, some of which emit most of their energy in spectacular bursts of TeV-photons\cite{michelson}. The blazar Markarian 421, at a redshift of $z=0.031$, has been observed to flare to a TeV flux exceeding the emission of the Crab, the most powerful galactic TeV source, by over one order of magnitudes. Some flares are as short as 15\,minutes in duration. Although correlated emission in X-rays and TeV gamma rays strongly favors synchrotron radiation by shock-accelerated electrons followed by inverse Compton scattering, the extreme behavior of the sources necessitates boost factors of order 30 to accommodate the observations.

But there is also the dog that didn't bark. No conclusive evidence has emerged for $\pi^0$ origin of the TeV radiation and, therefore, no cosmic ray sources have been identified. As previously discussed, recent observations of unusual gamma ray emission by molecular clouds may have produced the first indirect evidence for accelerated beams.

It has been realized for some time that the TeV sky is obscured by infrared light\cite{steckerchalonge}. Peak absorption of a photon of energy $E_{\gamma}$ by the reaction $\gamma(E_{\gamma}) + \gamma_{bkg} \rightarrow e^+ + e^-$ is by background photons of wavelength
\begin{equation}
\lambda_{bkg}(\mu m) = 1.24\,E_{\gamma}\,\rm(TeV) \,.
\end{equation}
Here the wavelength corresponds to the peak in the pair production cross section. TeV photons are absorbed on infrared background light emitted by stars and dust. The dust absorbs visible and UV light that is reemitted at infrared wavelengths. The two sources show up as separate peaks in the universal background in the vicinity of 1\,$\mu$\,m and 100\,$\mu$\,m, respectively. While very interesting\cite{cesarsky}, this background is poorly understood and mapping diffuse infrared light presents TeV astronomy with a great opportunity. Performing such measurements is not straightforward because one must distinguish whether the spectral cutoff observed at tens of TeV, for instance in the spectra of the Markarian 501 and 421 sources, reflects the maximum energy of the accelerator or absorption in intergalactic space. Having in mind the conventional synchrotron/inverse Compton model of the flux, the fact that both sources have a qualitatively similar TeV cutoff and very different X-ray spectra has been interpreted as evidence for absorption on infrared background light.
   
Recently, both HEGRA\cite{hegra1426} and Whipple\cite{whipple1426} have detected a more distant blazar H1426+428 at a redshift of $z=0.129$, to be compared with $\simeq 0.03$ for both Markarian sources. The HEGRA spectrum suggests absorption on infrared light: inclusion of absorption improves by a factor 4 the $\chi^2$ per degree of freedom of a power-law fit to the observed spectrum; see Fig.\,8a. The data may even suggest a feature corresponding to the separate contributions of stars and dust to the background spectrum; see Fig.\,8b. This interpretation requires a TeV flux at the source that is an order of magnitude larger than the one observed, a possible challenge to the conventional interpretation in terms of inverse Compton scattering. Some day, one may contemplate mapping space-time by observing sources over a range of redshifts\cite{primack}.

%% fig8
\def\thefigure{\arabic{figure}a}
\begin{figure}
\centering\leavevmode
\includegraphics[width=3.25in]{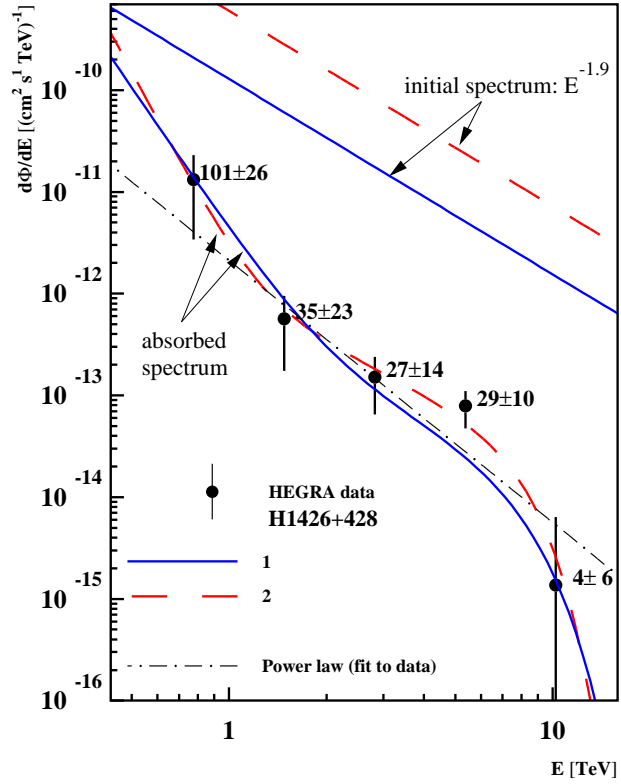}
\caption[]{HEGRA observations of TeV photons from the active galaxy H1426+428. Fits assuming that the initial flux has been partially absorbed by the universal background of infrared light (solid and dashed line) are compared to a powerlaw  fit to the data.}
\end{figure}

\addtocounter{figure}{-1}

\def\thefigure{\arabic{figure}b}

\begin{figure}
\centering\leavevmode
\includegraphics[width=3.5in]{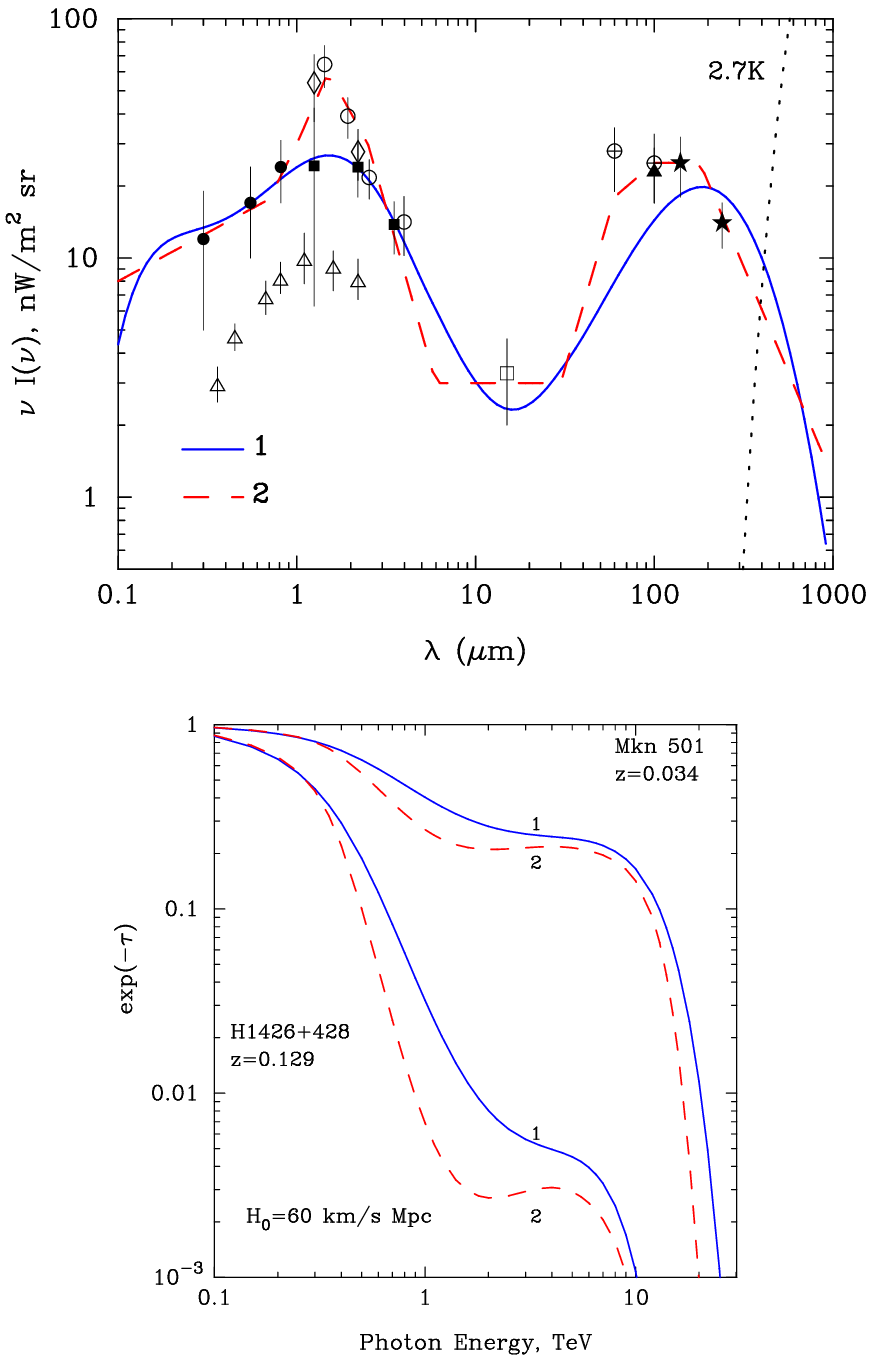}
\caption[]{The bottom panel shows the distortion of the observed spectrum of TeV photons resulting from absorption on infrared light of a powerlaw spectrum produced by a source at redshifts 0.034 and 0.129, respectively. Absorption has been evaluated for two fits to the universal spectrum of infrared photons; these are compared to data in the top panel.}
\end{figure}

\def\thefigure{\arabic{figure}}

The field of gamma ray astronomy clearly shows great promise and is buzzing with activity. Space-based detectors are extending their reach from GeV to TeV energy with AMS and, especially, GLAST\cite{michelson}, while the ground-based Cerenkov collaborations are designing second-generation instruments with lower thresholds. In the not so far future both techniques should produce overlapping measurements in the $10 {\sim} 10^2$~GeV energy range. All ground-based air Cerenkov experiments aim at lower threshold, better angular- and energy-resolution, and a longer duty cycle. One can however identify three pathways to reach these goals:

\begin{enumerate}

\item
larger mirror area, exploiting the parasitic use of solar collectors
during nighttime (CELESTE, STACEY and SOLAR\,Two),\cite{pare}
\item
better imaging of the Cerenkov footprint onto the 17\,m MAGIC mirror,\cite{magic}
\item
larger field of view and superior angular resolution and larger collection area using multiple telescopes (VERITAS, HEGRA and HESS).

\end{enumerate}

There is a dark horse in this race: Milagro.\cite{milagro} The instrument has been designed to lower the threshold of conventional air shower arrays to 100~GeV by instrumenting a pond of five million gallons of ultra-pure water with photomultipliers. For time-varying signals, such as bursts, the threshold may be lower. Milagro has observed the Crab and Markarian 421 and is presently constructing outrigger detectors and implementing smarter triggers to further lower the threshold.

\section{Neutrinos}

Whereas it has been realized for many decades that the case for neutrino astronomy is compelling\cite{nuast}, the real challenge has been to develop a reliable, expandable and affordable detector technology to build the kilometer-scale telescopes required to do the science. Suggestions to use a large volume of deep ocean water for high-energy neutrino astronomy were made as early as 1960. In the case of a high energy muon neutrino, for instance, the neutrino interacts with a hydrogen or oxygen nucleus in the water and produces a muon travelling in nearly the same direction as the neutrino. The blue Cerenkov light emitted along the muon's kilometer-long trajectory is detected by strings of photomultiplier tubes deployed deep below the surface. Collecting muons of neutrino origin far outside the detector, the effective detector volume exceeds the volume instrumented. With the first observation of neutrinos in the Lake Baikal\cite{baikal2,baikal3} and the (under-ice) South Pole neutrino telescopes\cite{nature}, there is optimism that the technological challenges to build kilometer-scale neutrino telescopes can finally be met.

The Baikal experiment represents a proof of concept for much larger deep ocean projects. These do however have the advantage of larger depth and optically superior water. Their real challenge is to find solutions to a variety of technological challenges for deploying a deep underwater detector. The European collaborations ANTARES\cite{antares,antares1,antares2} and
NESTOR\cite{nestor,nestor1} have planned initial deployments of large-area detectors in the Mediterranean Sea in the near future. The Italian NEMO Collaboration is conducting site studies and R\&D for a future kilometer-scale detector in the Mediterranean\cite{NEMO}.

\begin{figure}[h]
\centering\leavevmode
\includegraphics[width=3in]{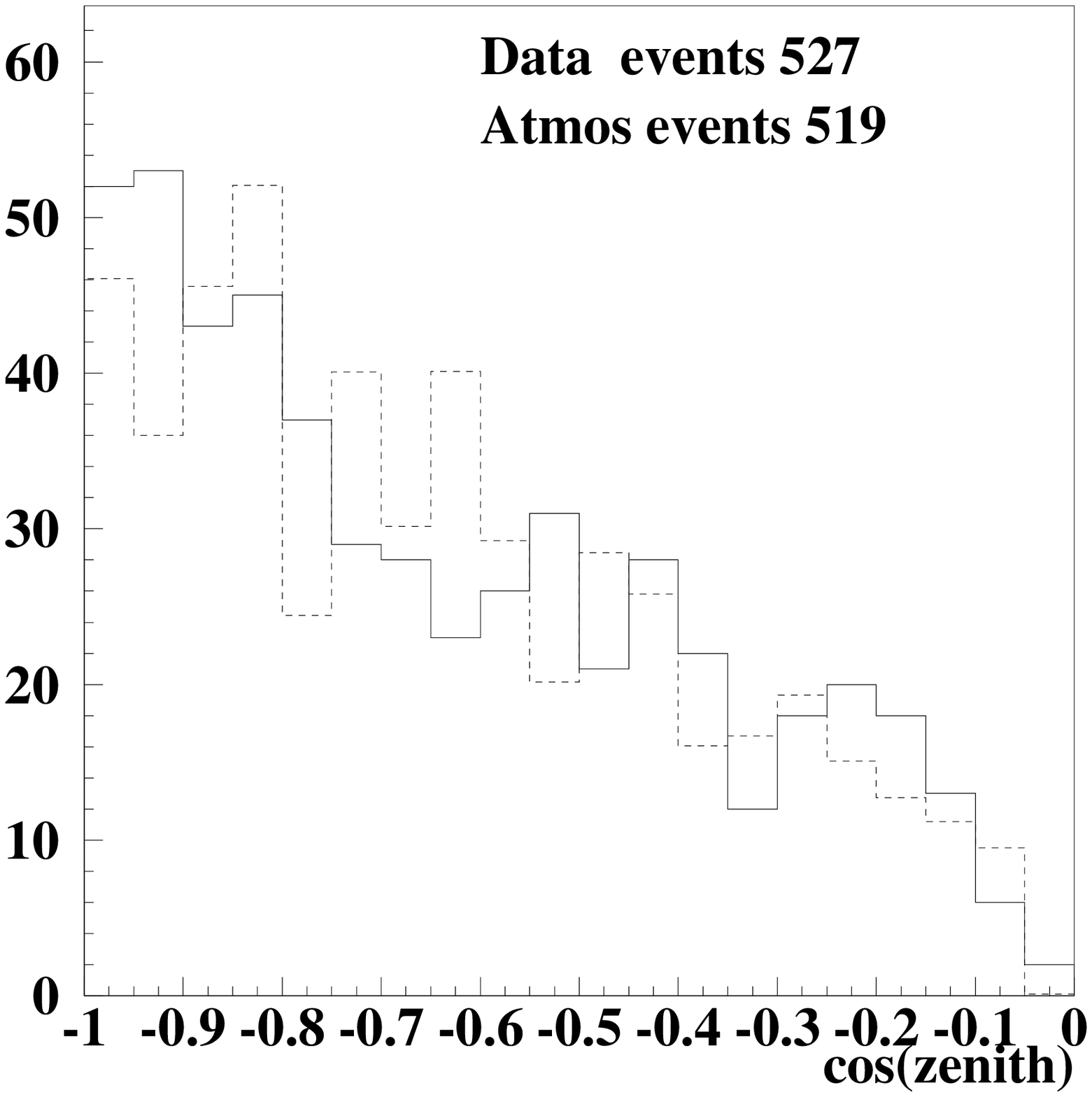}    %% old name {amanda-ii-atmnu.eps}
\caption[]{Number of upward-going muon events in AMANDA-II data from the
         year 2000 as a function of zenith angle, using a preliminary
         set of selection criteria.  There are a total of 527 events
         in the data or roughly 4 per day (solid line), while 519 events are predicted by the
         atmospheric neutrino Monte Carlo (dashed line).  Simulations
         indicate that these events have an energy of $\rm 60~GeV < E_\nu < 10~TeV$.  With more sophisticated selection criteria one expects larger event rates and improved response near the horizon.}
\label{fig:aii-atmnu}
\end{figure}

It is however the AMANDA detector using natural Antarctic ice as a Cerenkov detector that has already achieved the $> 10^4$\,m$^2$ telescope area envisaged by the DUMAND project a quarter of a century ago\cite{dumand}. It has operated for 3 years with 302 optical sensors and for another 3 years with 677. More than 3000 neutrinos, well separated from background, have been collected. The detector's performance has been calibrated by reconstructing upward-going muons produced by atmospheric muon neutrinos~\cite{nature,b10-atmnu}. A  preliminary analysis of atmospheric neutrino data taken with the completed AMANDA-II detector in 2000 is shown in Figure~\ref{fig:aii-atmnu}. After reconstruction of all muon tracks, collected at a rate of approximately 100 per second, two quality cuts are sufficient to separate atmospheric neutrinos from the background of cosmic ray muons that are more numerous by a factor $10^7$. The agreement between data and simulations, based on extrapolation of the Superkamiokande data to TeV energies, is excellent. The cuts simply require; i) a high quality fit and ii) a uniform distribution of the light along the muon track as is expected for relatively low energy muons tracks initiated by low energy atmospheric neutrinos.  Because of the simplified and robust analysis, made possible by the excellent coverage of muon tracks by the large AMANDA II detector, neutrino separation from background is now possible in real time and has been implemented since early 2002. The rate is 4 neutrinos per day; an event is shown in Fig.\,10.

%Figure 9

%% fig 10
\begin{figure}[t]
\centering\leavevmode
\includegraphics[width=2.25in]{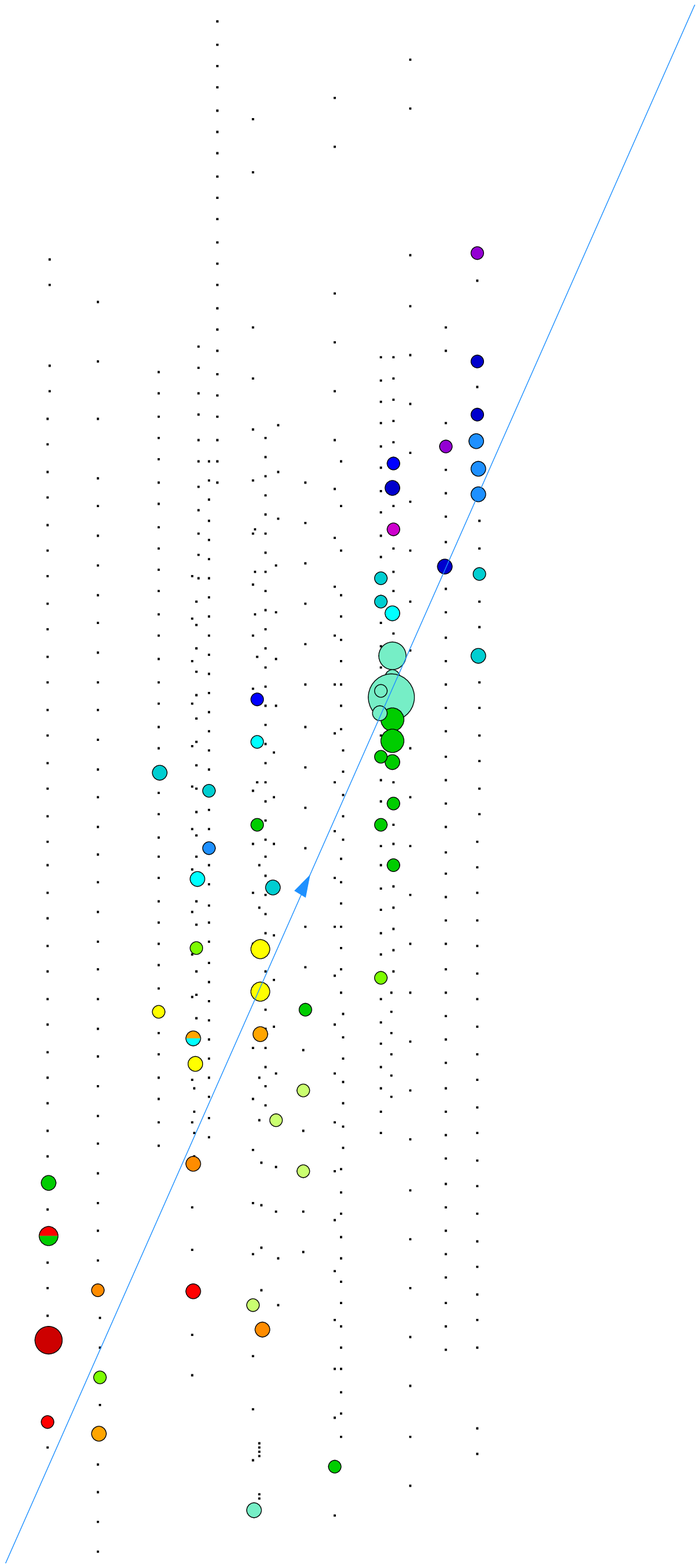}
\caption[]{Upward muon track in the completed AMANDA II detector. Black dots represent optical modules while the circles represent the time (shading) and amplitude (size) of the light registered by photomultipliers reporting signals associated with the fitted track.}
\end{figure}

Using the first of 3 years of AMANDA II data, the collaboration is performing a general search for  the continuous emission of muon neutrinos from a spatially localized direction in the northern sky\cite{nu2002}. Backgrounds are reduced by requiring a statistically significant enhancement in the number of reconstructed upward-going muons in a small bin in solid angle. The background for a particular bin can be calculated from the data by averaging over the data outside the bin in the same declination band. Preliminary sensitivities to a sample of point sources are given in Table~\ref{table:point_source_sensitivities}. In order to achieve blindness in this analysis the right ascension of each event (i.e., its azimuthal angle) has been scrambled. At the South Pole this effectively scrambles the event time. The data will only be unscrambled after final selection criteria have been set. The (scrambled) skyplot is shown in Fig.\,11. AMANDA has reached an interesting milestone: after unblinding the data, sources with a $E^{-2}$ spectrum should be observed provided the number of gamma rays and neutrinos are roughly equal as expected from cosmic ray accelerators producing pions\cite{alvarezhalzen}.

\begin{table}
\tbl{Example of AMANDA-II sensitivity to point sources in data taken in 2000.  The sensitivity
         is defined as the predicted average limit from an
         ensemble of experiments with no signal, and is calculated using
         background levels predicted from off-source data.\vspace*{1pt}}
{\footnotesize
\begin{tabular}{|lr@{\qquad}r@{\qquad\qquad}r@{\qquad\qquad}|} \hline
Source              & \multicolumn{1}{c}{Declination}
                            & \multicolumn{1}{c}{$\mu$ ($\times 10^
{-15}$cm$^{-2}$s$^{-1}$)}
                                   & \multicolumn{1}{c|}{$\nu$ ($\times 10^{-8}$cm$^{-2}$s$^{-1}$)}\\ \hline
%%%Markarian 421       &       & 2.6  & 1.1 \\
SS433               & 5.0   & 11.0 & 2.4 \\
Crab                & 22.0  & 4.0  & 1.3 \\
Markarian 501       & 39.8  & 2.5  & 1.0 \\
Cygnus X-3          & 41.5  & 2.6  & 1.1 \\
Cass. A             & 58.8  & 2.1  & 1.0 \\  \hline
\end{tabular}}
\label{table:point_source_sensitivities}
\end{table}

%% fig. 11
\begin{figure}
\centering\leavevmode
\includegraphics[width=4.25in]{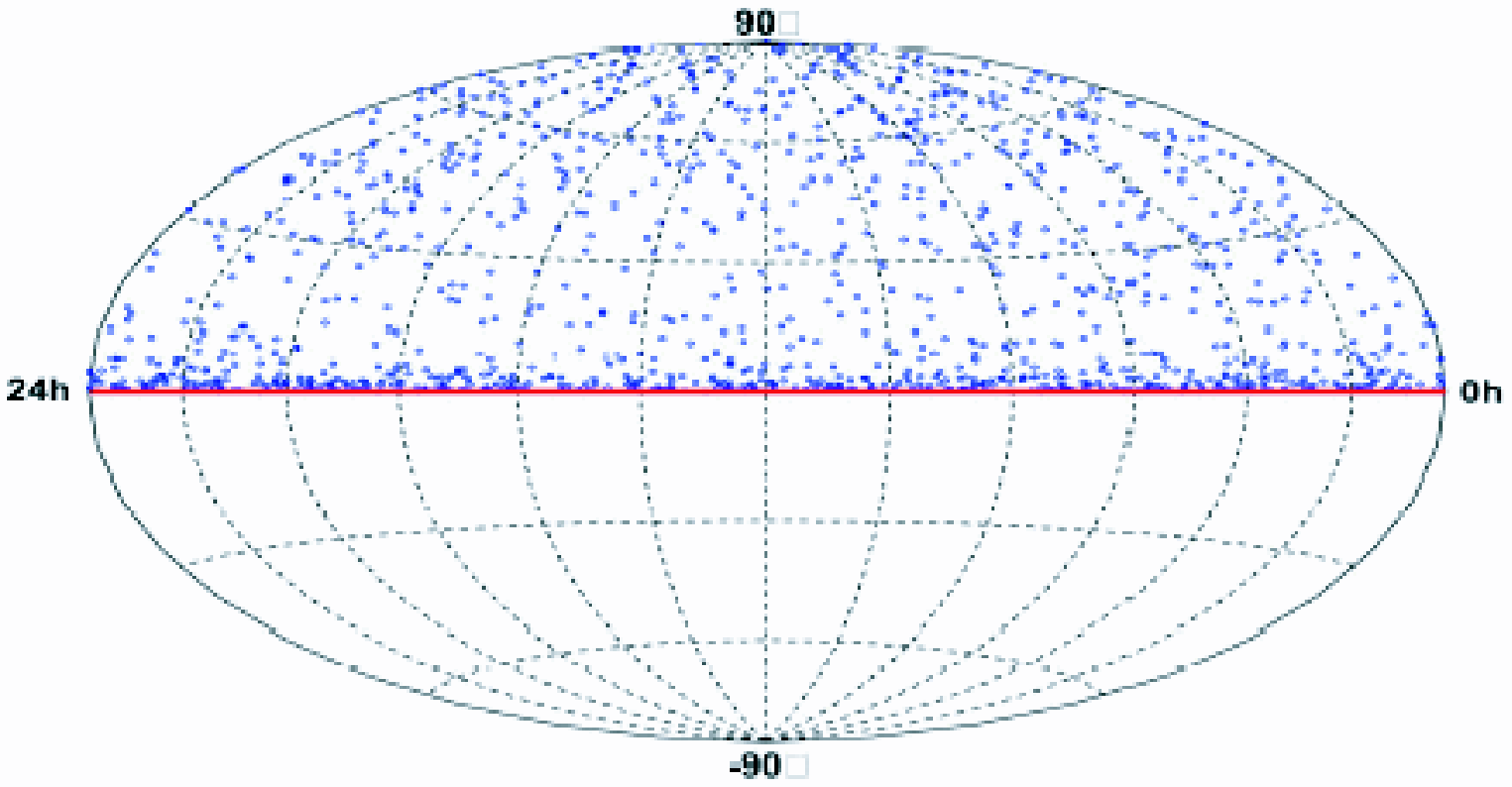}
\caption[]{Skymap showing declination (blinded) and right ascension of neutrinos detected by the completed AMANDA-II detector during its first Antarctic winter of operation in 2000. Estimated sensitivities to various point sources are listed in the table.}
\end{figure}

The AMANDA II search for a diffuse flux from sources of high energy $\nu_\mu$--induced muons has reached a {\it sensitivity} of 
$1.5\times10^{-7}\rm\,GeV\ cm^{-2}\,s^{-1}\,sr^{-1}$, well below the flux predicted from active galaxies\cite{steckerchalonge}. When unblinded, the data should reveal a high energy flux from active galaxies or, alternatively, significantly constrain the fraction of energy going into high energy protons, {\it i.e.} cosmic rays. A search for $\nu_\mu$--induced muons from gamma-ray bursts is in progress. It leverages temporal and directional information from satellite observations to realize a nearly background-free analysis. Data spanning the years 1997--2000 have been analyzed from some 500 GRB in coincidence with satellite experiments.

Overall, AMANDA represents a proof of concept for the construction of the kilometer-scale neutrino observatory, IceCube\cite{pdd}, now under construction. IceCube is an instrument optimized to detect and characterize neutrinos of all flavors from sub-TeV to the highest energies; see Fig.\,12. It will consist of 80 kilometer-length strings, each instrumented with 60 10-inch photomultipliers spaced 17\,m apart. The deepest module is 2.4~km below the surface. The strings are arranged at the apexes of equilateral triangles 125\,m on a side. The instrumented detector volume is a cubic kilometer; the detector will therefore exceed a kilometer square in effective telescope area for all flavors. A surface air shower detector, IceTop, consisting of 160 Auger-style Cerenkov detectors deployed over 1\,km$^{2}$ above IceCube, augments the deep ice component by providing a tool for calibration, background rejection and air shower physics.

% figure 12
\begin{figure}[h]
\centering\leavevmode
\includegraphics[width=2.5in]{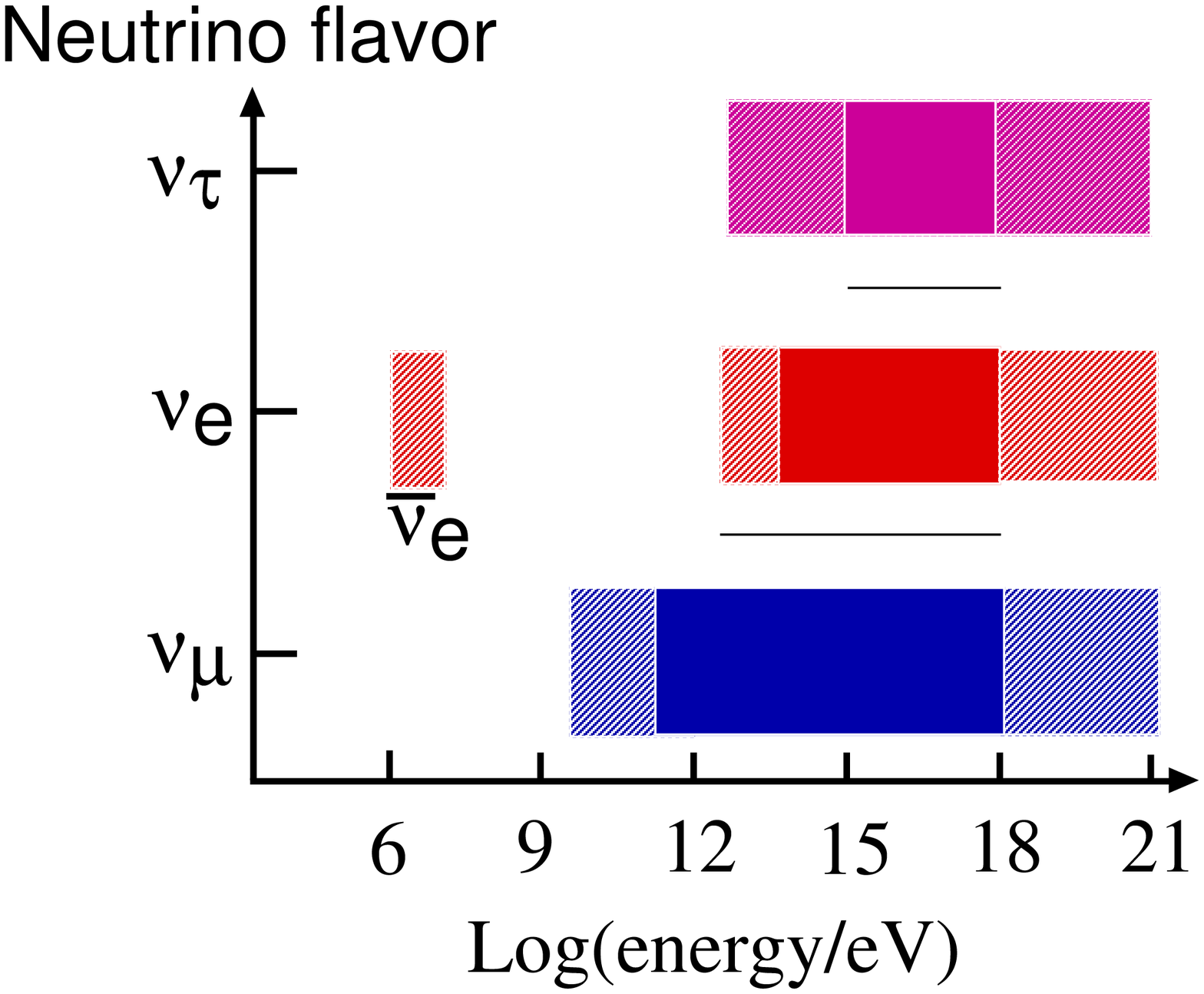}
\caption[]{Although IceCube detects neutrinos of all energies and flavor above a
threshold of $\sim 0.1$\,TeV, it can also identify their flavor and measure
their energy only in the ranges shown.}
\end{figure}

IceCube will offer great advantages over AMANDA beyond its larger size: it will have a higher efficiency and superior angular resolution in reconstructing tracks, map showers from electron- and tau-neutrinos (events where both the production and decay of a $\tau$ produced by a $\nu_{\tau}$ can be identified) and, most importantly, measure neutrino energy. Simulations, verified using AMANDA data, indicate that the direction of muons can be determined with sub-degree accuracy and their energy measured to better than 30\% in the logarithm of the energy. The direction of showers will be reconstructed to better than 10$^\circ$ above 10\,TeV and the response in energy is linear and better than 20\%. Energy resolution is critical because, once one establishes that the energy exceeds 1\,PeV, there is no atmospheric muon or neutrino background in a kilometer-square detector and full sky coverage is achieved, including the galactic center\cite{pdd}.

At lower energies the backgrounds are down-going cosmic ray muons, atmospheric neutrinos, and the dark noise signals produced by the photomultipliers themselves.  The simulated trigger  rate of down-going cosmic ray muons in IceCube is 1700\,Hz while the rate of atmospheric neutrinos ($\nu_\mu$ and $\overline{\nu}_\mu$) at trigger level is 300 per day.  Depending on the type of signal to be searched for, this background is rejected using direction, energy, and neutrino flavor. 

It is the hope that within 5 years IceCube should reach the kilometer-scale required to do the science which, beyond the detection of neutrinos associated with cosmic rays, also includes the search for dark matter and the study of neutrino physics at energies far beyond the reach of accelerators.

Recently, a wide array of projects have been initiated to detect neutrinos of the highest energies, typically above a threshold of 10 EeV, exploring other experimental signatures: horizontal air showers and acoustic or radio emission from neutrino-induced showers. Some of these experiments, such as the Radio Ice Cerenkov Experiment\cite{frichter} and an acoustic array in the Caribbean\cite{lehtinen}, have taken data; others are under construction, such as the Antarctic Impulsive Transient Antenna\cite{gorham}.  

\section*{Acknowledgements}
I gratefully acknowledge Felix Aharonian, Tom Gaisser, Frank Krennich, Pierre Sokolsky and Masahiro Teshima for contributed material. I thank M.~Stamatikos for reading the manuscript. This research was supported in part by the National Science Foundation under Grant No.~OPP-0236449, in part by the U.S.~Department of Energy under Grant No.~DE-FG02-95ER40896, and in part by the University of Wisconsin Research Committee with funds granted by the Wisconsin Alumni Research Foundation.

\end{document}